**Motion and temporal $B_0$ shift corrections for quantitative susceptibility mapping (QSM) and $R_2$\* mapping using dual-echo spiral navigators and conjugate-phase reconstruction**

Yuguang Meng[1], Jason W. Allen[1,2,4], Vahid Khalilzad Sharghi[3], Deqiang Qiu[1,4*]

[1]Department of Radiology and Imaging Sciences, Emory University, Atlanta, Georgia, USA

[2]Department of Radiology and Imaging Sciences, Indiana University, Indianapolis, Indiana, USA

[3]Siemens Heathineers, Atlanta, Georgia, USA

[4]Department of Biomedical Engineering, Emory University and Georgia Institute of Technology, Atlanta, Georgia, USA

[*]Correspondence to:    Deqiang Qiu, PhD

Department of Radiology and Imaging Sciences,

Emory University, Atlanta, GA 30322, USA.

Email: deqiang.qiu@emory.edu

Figure(s):    7

Total words:    4308

**ACKNOWLEDGEMENTS**

The study was supported by the National Institutes of Health Grant/Award Numbers: R21AG064405, R01AG07093, R01AG072603, P30AG066511.

Submitted to ***Magnetic Resonance in Medicine***

## Abstract

**Purpose:**    To develop an efficient navigator-based motion and temporal $B_0$ shift correction technique for 3D multi-echo gradient-echo (ME-GRE) MRI for quantitative susceptibility mapping (QSM) and $R_2$* mapping.

**Theory and Methods:**    A dual-echo 3D spiral navigator was designed to interleave with the Cartesian ME-GRE acquisitions, allowing the acquisition of both low- and high-echo time signals. We additionally designed a novel conjugate-phase based reconstruction method for the joint correction of motion and temporal $B_0$ shifts. We performed both numerical simulation and *in vivo* human scans to assess the performance of the methods.

**Results:**    Numerical simulation and human brain scans demonstrated that the proposed technique successfully corrected artifacts induced by both head motions and temporal $B_0$ changes. Efficient $B_0$-change correction with conjugate-phase reconstruction can be performed on less than 10 clustered k-space segments. In vivo scans showed that combining temporal $B_0$ correction with motion correction further reduced artifacts and improved image quality in both $R_2^*$ and QSM images.

**Conclusion:**    Our proposed approach of using 3D spiral navigators and a novel conjugate-phase reconstruction method can improve susceptibility-related measurements using MR.

**Introduction**

MR scans are sensitive to motion (1,2), and many methods have been developed to minimize artifacts due to subject motion (2). Compared to spin-echo, gradient-echo (GRE) based imaging methods such as $T_2^*$, susceptibility-weighted imaging (SWI) or quantitative susceptibility mapping (QSM) generally require longer echo-time (TE) to achieve adequate image contrast between tissues. It has been shown that these GRE-based MRI techniques are useful for the study of neurological diseases such as Alzheimer's Disease and multiple sclerosis (3,4). However, the GRE images are more susceptible to movement because they are additionally sensitive to magnetic field ($B_0$) variations due to subject motion, respiration, and cardiac pulsation (5-12). As a result, the consideration and correction for temporal $B_0$ variations are critical in improving the image quality of GRE images and their derived images such as $R_2^*$ and QSM images. While previous efforts have developed methods to perform motion and $B_0$-variation corrections on $T_2^*$ images, limited work has been done to evaluate their effectiveness on QSM, which is particularly sensitive to phase variations (5-13). Computationally efficient algorithms are also needed to translate these methods to the clinic.

Motion correction techniques in MR can be broadly categorized to MR-navigator based approaches (either self-navigated or with acquisition of additional navigators) (5-9,11,14,15), and those based on external motion-monitoring devices, such as optical cameras (13,16-18), or the combination of MR-navigator and external motion-monitoring device (10). While optical camera-based approaches have found applications in many MR methods, they require additional devices and cannot directly monitor the temporal $B_0$ field variations (10,13,16-18). On the other hand, despite the cost of lengthening scanner time and lower temporal resolution in motion-tracking, navigator-based motion correction approaches do not require additional devices (5-9,11,14,15), allowing wider adoption. Navigator-based MRI sequences for motion correction have been applied to high-resolution $T_1$-weighted imaging (MPRAGE) and $R_2^*/T_2^*$ imaging (5,6,11,14). In these designs, low-resolution volumetric navigators are inserted between the k-space lines of the host sequence. With the application of parallel imaging, these navigators can provide sufficient temporal and spatial resolution for the determination of motion that can be incorporated in the image reconstruction (5,6,9,11,14). Besides estimating motion, navigator-based motion correction methods have also been developed to correct for the temporal $B_0$ variations in $R_2^*/T_2^*$ weighted imaging (5-11). For example, Gretsch et al (6) used double-echo short TR 3D GRE acquisitions of the selectively excited fat signal as navigators to reduce the acquisition time compared to the volumetric navigators, and used these to correct motion and $B_0$ shift artefacts in $T_2^*$ imaging. However, due to the use of a fat-selective excitation, there were inevitable disruptions to the steady-state of water signals due to the insertion of the navigators in these approaches, and the whole brain $B_0$ field variations could not be sufficiently obtained from the scalp lipid signals. In another approach (5,11), navigators with segmented echo-planar imaging (EPI) were firstly acquired as the early echoes with the acquisition of later echoes for the host 3D GRE in each RF excitation. This approach has the advantage of higher temporal resolution for the navigators and the water signal maintains the same steady state. However, this design prevents the acquisition of early echoes with shorter echo times for 3D GRE, and these early echoes are important for the estimation of pseudo proton density at zero echo time and potentially $T_2^*/R_2^*$ and QSM maps as well as for multi-compartment modeling (11,19). Furthermore, despite these efforts on motion and temporal $B_0$-variation corrections for $T_{2w}^*$ imaging and $T_2^*/R_2^*$ mapping as well as motion correction but without $B_0$-variation correction for QSM, the combined effects of motion and temporal $B_0$ variations on QSM have not been well studied (5-11,13). Since QSM utilizes the phase information of the acquired signal, $B_0$ variations might result in different effects on QSM than magnitude images (20). Additionally, previous joint motion and $B_0$ variations correction methods for $T_2^*/R_2^*$ mapping used algorithms that are not easily parallelizable, limiting their potential for clinical translation.

In this work, we aimed to develop an efficient, navigator-based multi-echo GRE (ME-GRE) MRI with motion and $B_0$-variation corrections for both quantitative $R_2^*$ measurement and QSM. In our MRI pulse sequence design, dual-echo 3D spiral navigators are interleaved with Cartesian ME-GRE signal excitations and acquisitions, which facilitates echo time adjustments as in conventional gradient-echo MRI for quantitative $R_2^*$ measurement and/or QSM. Compared to EPI-based navigators that were used in previous studies, we designed a dual-echo 3D spiral imaging-based navigator. Spiral imaging has been shown to have more efficient k-space coverage compared to EPI (21). Dual-echo 3D spiral acquisition also allows very short TE readout in the first echo, which helps to reduce signal loss due to $T_2^*$ decay, while the second echo provides sensitivity to $B_0$ variations. For image reconstruction, a novel conjugate-phase reconstruction approach is proposed to correct the temporal $B_0$ field variations during the entire 3D k-space acquisition period. The reconstruction is performed by dividing the procedure into motion correction followed by temporal $B_0$-variation corrections on clustered k-space segments with similar $B_0$ changes. The proposed approach can facilitate parallel reconstruction. As compared to the previous correction methods (5,6), the proposed reconstruction is simple and intuitive, and does not need to modify the original motion correction algorithm, which allows the retention of a variety of motion correction methods by attaching the $B_0$-variation correction as an additional step. The feasibility and performance are demonstrated in $T_2^*$-weighed imaging, $R_2^*$ measurement and QSM at 3.0 T.

**Methods**

*Theory*

With subject motion parameterized by translation matrix $T$ and rotation matrix $R$, the acquired signal $S(\boldsymbol{k}, c)$ for an intended k-space location $\boldsymbol{k}$ and coil element $c$ is:

$$S(\boldsymbol{k}, c) = e^{-i\boldsymbol{k}\cdot\boldsymbol{T}} \cdot \int \rho(\boldsymbol{r}) B_{1,c}(\boldsymbol{r}) e^{-i\gamma\Delta B_{0,k}(r)\cdot TE} e^{-iR'\cdot\boldsymbol{k}\cdot\boldsymbol{r}} \cdot d\boldsymbol{r}, \qquad [1]$$

where $\gamma$ is the gyromagnetic ratio, $\boldsymbol{r}$ is spatial coordinate in the patient coordinate frame, $\rho(\boldsymbol{r})$ is the spatial distribution of the magnetization density that we would like to recover and encapsulates the effects of $T_1$ and $T_2^*$ relaxation among others, $B_{1,c}(\boldsymbol{r})$ is the receive coil sensitivity map for the $c$-th coil element, $R'$ is the transpose of $R$ that be measured from the 3D navigators during acquisition, TE is the echo time in the ME-GRE train, $\Delta B_{0,k}(r)$ is the spatial $B_0$ field variation during the acquisition of k-space ($\boldsymbol{k}$). In this way, the induced phase variation during acquisition is accumulated by $\Delta B_{0,k}(\boldsymbol{r})$ at TE.

In the presence of translational and rotational movements, there is k-space rotation and linear phase accumulation. Let $S'(\boldsymbol{k}', c)$ denotes the k-space for $c$-th coil element without the contamination of motion. Using Eq. [1] and setting $\boldsymbol{k}' = R'\cdot\boldsymbol{k}$, we have:

$$S'(\boldsymbol{k}', c) = S(\boldsymbol{k}, c) \cdot e^{i\boldsymbol{k}\cdot\boldsymbol{T}} = \int \rho(\boldsymbol{r}) B_{1,c}(\boldsymbol{r}) e^{-i\gamma\Delta B_{0,k'}(r)\cdot TE} e^{-ik'\cdot\boldsymbol{r}} \cdot d\boldsymbol{r}, \qquad [2]$$

Mathematically, Eq. [2] can be rewritten in a discrete form as

$$S' = EM, \qquad [3]$$

where $M$ is a vector of the magnetization density with the entry $\rho(r)$ and the $E$ is a transformation matrix with the entry:

$$E_{(k',c),r} = B_{1,c}(r) e^{-i\gamma\Delta B_{0,k'}(r)\cdot TE} e^{-ik'\cdot r}. \qquad [4]$$

The pseudo-inverse solution of $M$ is:

$$\widehat{M} = (E^\dagger E)^{-1} E^\dagger S', \qquad [5]$$

where $E^\dagger$ is the Hermitian conjugate of $E$.

Assuming that there are small variations in $\Delta B_{0,k}(r)$ with respect to location, the term $E^\dagger E$ can be approximated as $F^\dagger F$ in which $F^\dagger$ and $F$ are the Fourier transformation and inverse Fourier transformation matrices with receive coil sensitivity respectively (see Appendix), and their entries are:

$$F_{(k',c),r} = B_{1,c}(r) \cdot e^{-ik' \cdot r},$$ [6]

$$F_{r,(k',c)}^{\dagger} = B_{1,c}(r) \cdot e^{ik' \cdot r}.$$ [7]

Each k-space data acquired at a specific k-space location may experience a unique $B_0$ drift. If the k-space data in each $j$-th k-space segment ($1 \leq j \leq n$) are undergoing the same $\Delta B_{0,j}$ during acquisition, Eq. [5] can be solved with segmented k-space reconstruction and rewritten as

$$\hat{M} = \sum_{j=1}^{n} B_j^{\dagger}[(F^+F)^{-1}F^{\dagger}S_j'],$$ [8]

where $B_j^{\dagger}$ is the complex conjugate diagonal matrix in which the element at location $(r, r)$ is:

$$B_j^{\dagger}(r,r) = e^{i\gamma \Delta B_{0,j}(r) \cdot TE}.$$ [9]

The reconstruction flow-chart was shown in Fig.1. The image was firstly reconstructed from $n$ k-space partitions with motion correction with the translational and rotational information, in which partial inverse Fourier transformation on the parallelly-imaged data is solved by a linear conjugate gradient algorithm. During the $\Delta B_0$ correction step, similar to the off-resonance correction method where the signal acquisition window was divided into segments to approximate small phase accumulation during the segments, the conjugate-phase reconstruction is used here to correct the phase accrual difference due to the difference of $\Delta B_0$ among the divided $n$ 3D k-space segments in which the clustered k-space lines are undergoing similar $\Delta B_0$ during acquisition (22). The motion-corrected results for each k-space segment were multiplied by $\exp[i\phi_j(TE)]$ where $\phi_j = \gamma \cdot \Delta B_{0,j}(r) \cdot TE$ ($1 \leq j \leq n$) is the conjugate-phase accrual for segment $j$ at TE. Since the k-space segmental reconstruction is independent of each other, the reconstruction procedure can be performed in parallel. Finally, all reconstructed results after the conjugate-phase corrections are summed together to make the final image as with Eq. [8].

*Numerical simulation*

As the goal is to evaluate the effects of motion and $B_0$ field change on different reconstructions, a simulated 2D Shepp-Logan head phantom was used to evaluate the effectiveness of different correction methods. The magnitude of the 2D phantom was simulated with matrix size 320×320 with the definitions of horizontal ($x$) and vertical ($y$) directions (left in Fig. 2a). The simulated $B_0$ field change map contains linear and non-linear $B_0$ field change, i.e., $\Delta B_0 = \beta_0 + \beta_x \cdot x + \beta_y \cdot y + \beta_{nl}(x, y)$, where $\beta_0$ is the constant term, and $\beta_x$ and $\beta_y$ are the first-order terms, and $\beta_{nl}(x, y)$ is the spatially dependent (i.e., $x$ and $y$) non-linear higher-order (2nd and above) of the $B_0$ field change approximation. During the acquisition of the central horizontal 10 k-space lines, motion was simulated with 2-pixel translation in the horizontal ($x$) direction and 2º in-plain rotation, making $\Delta B_0$ containing -0.01 Hz/voxel first-order term in the $y$-direction across the 2D phantom (Fig. 2a-b). The k-space data were simulated with simulated 32-channel receiver sensitivity maps as in Eq. [1] and TE = 32 ms, R = 2 in phase-encoding direction with 24 reference lines. Zero-mean complex Gaussian-distributed noise was added to the simulated k-space data with SNR 5 at the highest frequency. A sensitivity encoding (SENSE) reconstruction method with 10 conjugate-gradient iterations provided by BART Toolbox was performed (23). For comparison, the simulated phantom was reconstructed without any correction (uncorrected), with motion correction only (MC), joint motion and linear $\Delta B_0$ corrections (MC+Lin.$\Delta B_0$) (6), and joint motion and $\Delta B_0$ corrections using conjugate-phase information (MC+Conj.$\Delta B_0$). Root-mean-square-error (RMSE) between the reconstructed images and the simulated phantom (the ground truth) were evaluated in terms of the magnitude and the phase respectively.

*MRI pulse sequence design and imaging*

The 3D ME-GRE sequence with navigators for motion and $B_0$ shift tracking was implemented on a Siemens 3T MRI scanner (Prisma model, Siemens Healthineers) with a 32-channel head receive coils and IDEA software (version: VE11C, Siemens Healthineers). As shown in Fig. 3, dual-echo 3D volumetric stack-of-spirals navigators were interleaved with GRE acquisitions (24). The following parameters were used for the 3D volumetric spiral navigators: single-shot for each slice-selective $k_z$ plane with readout duration = 16.1 ms, matrix size = 48×48×32, isotropic resolution = 4.6 mm, acceleration factor R = 4 in the slice-selective direction, dual gradient-recall echoes with TEs = 2.4 and 19 ms. Each under-sampled spiral navigator acquisition was interleaved with 25 executions of the GRE base sequence. The host Cartesian ME-GRE acquisitions were performed with 5 echoes, first TE = 4 ms and echo-spacing = 7 ms, matrix size = 320×320×104, acceleration factor R = 2×2 in the phase- and slice-encoding directions with 24 reference lines, and resolution = 0.7 mm × 0.7 mm × 1.4 mm. The same RF flip angle of 15º and TR of 40 ms were used for both the spiral navigators and the ME-GRE excitations to maintain a steady state for the signal. Before the 3D navigator acquisitions, 100 dummy scans were performed to achieve the steady-state signal followed by a fully-sampled 3D spiral signal with minimal TE for receiver coil sensitivity map calibration in the navigator reconstructions. The 3D human brain scans for six normal subjects were performed with written consent under the approval from the Emory University Institutional Review Board. The scanning time was around 8.5 minutes for each subject.

*Volumetric 3D spiral navigator reconstruction and processing*

The 3D spiral navigators were reconstructed by ESPIRiT algorithm implemented as the Berkeley Advanced Reconstruction Toolbox (BART) toolbox (23). The rigid-body translations and rotations relative to the reference position/posture (i.e., the reference) that took place at the time close to the acquisition of the 3D GRE k-space center were estimated from the magnitude images of the spiral navigators using the "spm_realign" function in SPM12 toolbox (UCL, UK) with the following parameters: the 1.0 highest "quality" for selecting the optimized number of pixels to the estimation of the realignment parameters, 5 mm full-width half-maximum Gaussian smoothing, and 4 mm separation of image samples between the points sampled in the reference image. After registration to the reference volume of the navigators, the $B_0$ field change relative to the reference $B_0$ field were derived from the phase information with the TE interval of the navigator ($\Delta TE_{Nav}$ in Fig. 3), followed by phase unwrapping with Laplace algorithm (25). The unwrapped phase was then filtered with a 3D Gaussian filter of which the standard deviation (sigma) was one voxel and the kernel size was five voxels, and fitted with a 4th-order spherical harmonics approximation model.

*GRE reconstruction with motion and $\Delta B_0$ corrections*

In the image reconstruction procedure, the acquired k-space lines were divided into segments with similar $\Delta B_0$ during the acquisition of these k-space lines using a k-means clustering method. The more clusters are used, the more correction operations for $B_0$ shift will be performed at the cost of reconstruction time. Taking the maximum 20 segments as a reference for clustering the k-space lines, the RMSE of the $\Delta B_0$ relative to the centroids of n (1-20) clusters were calculated ($RMSE_n$). Then, the number of the cluster was chosen by searching for the minimum cluster number with which the absolute difference between $RMSE_1$ and $RMSE_n$ is less than 10% of the difference between $RMSE_1$ and $RMSE_{20}$. In this way, it is assumed that the chosen cluster number provides sufficient approximation for clustering the k-space lines with similar $\Delta B_0$ with the threshold (5). The reconstructions were performed on each k-space segment with translational and rotational motion corrections, by a sensitivity encoding (SENSE) reconstruction method with 10 conjugate-gradient iterations provided by BART Toolbox (23). Following the correction for the temporal $B_0$ changes for each k-space segment, a summation was performed across all k-space segments according to Eq. [8].

*QSM reconstruction*

The QSM reconstruction was performed with MEDI software (available from Cornell QSM toolbox). In the reconstruction, the brain mask was obtained using a fractional intensity threshold of 0.5 with BET2 (FSL toolbox developed by FMRIB, Oxford). Within the brain mask, the estimation of the frequency offset for each voxel in the images was performed using a complex fitting, followed by spatial phase unwrapping (26). Subsequently, background field removal was carried out (27). Using CSF mask for zero referencing, quantitative susceptibility was calculated with an $\ell$1-norm penalty-based morphology-enabled dipole inversion (MEDI) method with the default Lagrange multiplier of 1000 (26,28).

*Statistics on human scans*

The human scans data were reconstructed using different reconstruction methods including one without any correction (uncorrected), with MC only, MC+Lin.$\Delta B_0$ and MC+Conj.$\Delta B_0$. For each of these reconstructions, $R_2^*$ and QSM were calculated from the reconstructed $T_{2w}^*$ images. Taking the ones with MC+Conj.$\Delta B_0$ corrections as references, RMSE of the images reconstructed from the references and the ones with other methods were evaluated across six subjects. Afterwards, one-way ANOVA with the different reconstructions as a within-subject factor followed by post-hoc multiple comparisons was performed to test the differences between the reconstruction methods for motion and $B_0$ shift corrections. P-values less than 0.05 were considered statistically significant.

**Results**

*Numerical simulation*

From Fig. 2c-d, numerical simulation showed that the RMSE of the magnitude/phase of the reconstructed image with MC or MC+Lin.$\Delta B_0$ was smaller compared to the one without any correction, though the reconstructed images were visually similar. With MC+Conj.$\Delta B_0$, the RMSE reduced further, and the reconstructed results were very close to the grand truth. In addition, it was shown that using MC or MC+Lin.$\Delta B_0$ there were still significant phase errors in the corrected image, while the one with MC+Conj.$\Delta B_0$ had minimal phase error.

*In vivo human brain scans*

In human brain scans, the quality of the isotropic 4.6 mm low-resolution 3D-spiral navigators were shown in Fig. 4, where there was no obvious distortion in both magnitude and phase images and the phase varied slowly spatially within ±50 Hz. Based on the low-resolution magnitude and phase images, the movement amplitude and temporal $\Delta B_0$ was identified from 348 navigators across the acquisition.

Fig. 5a shows movement and $B_0$ variation in an example subject who had typical unintentional gradual movements with relatively small amplitudes corresponding to a maximum translation of 2 mm and the rotation within 2º, producing no more than 0.2 Hz temporal $\Delta B_0$ variations averaged across the whole brain. To explore the correction effectiveness with $\Delta B_0$ clustering number, normalized root-mean-square error (NRMSE) of $T_{2w}^*$ (TE=32 ms) images between the reconstructed results with n ($1 \leq n \leq 20$) clusters and the maximum 20 clusters were calculated from a subject with significant motion, and the NRMSEs were less than 0.1% when more than 5 clusters were used (see Fig. S1 in supplemental materials). Thus, taking the maximum 20 clusters as a reference, K-means clustering was performed on the temporal $\Delta B_0$ by estimating the $\Delta B_0$ clustering error $|RMSE_n - RMSE_{20}|/|RMSE_1 - RMSE_{20}|$ as described in the method. From Fig. 5b, the $\Delta B_0$ clustering error decreased to less than 0.1 when 9 or more clusters were used. The spatial $B_0$ change varied slowly to the edge of brain within ±10 Hz, and the clustered $\Delta B_0$ maps (i.e., clusters 4, 6 and 8) had more significant change with movements. The reconstructed results were compared in Fig. 5c. Without any correction, there were remarkably dark and blur artifacts in the vein-like region (yellow rectangle) with significant $\Delta B_0$ as in the clustered $\Delta B_0$ maps in Fig. 5b. With MC, there was little image quality improvement in the vein-like areas, and such situation was similar when MC+Lin.$\Delta B_0$ was applied. The reconstructed one from MC + Conj.$\Delta B_0$ was taken as the reference because visually it had minimal motion-induced artifacts, which was also validated by numerical simulations. In the quantitative analysis, $R_2^*$ or QSM variations still existed even though MC or MC+Lin.$\Delta B_0$ was used. For example, $R_2^*$ or QSM would vary 15-20 Hz or ppb when reconstructed with MC+Lin.$\Delta B_0$ compared with that from MC+Conj.$\Delta B_0$, as shown in the vein-like areas (Fig. 5c).

Fig. 6 shows results from another subject with an unintentional sudden and large movement during acquisition. In this example, there was little movement during most of the acquisition time except during the acquisition of outer k-space when there was around 5 mm translation and 4º rotation inducing around 0.5 Hz $B_0$ variations averaged across the whole brain. Using 7 clustered $\Delta B_0$ maps, $\Delta B_0$ was captured during the large movements (i.e., clusters 5 and 7). In contrast to the results with gradual movement with small amplitudes, $\Delta B_0$ showed significant changes across most part of the anterior brain, especially when there were larger movements captured in cluster 5 in Fig. 6b. Compared to the results from MC+Conj.$\Delta B_0$, other reconstructed results exhibited blurring in the subcortical area, even though MC+Lin.$\Delta B_0$ was used. The measured $R_2^*$ or QSM would vary 20-30 Hz or ppb when they were reconstructed without MC+Conj.$\Delta B_0$.

RMSE between reconstruction from MC+Conj.$\Delta B_0$ as the reference and the other reconstructions from six subjects are shown and compared in Fig. 7. Statistical analysis showed that there were significant differences between the reconstructions for $T_{2w}^*$ ($p < 0.001$), $R_2^*$ ($p = 0.01$) and QSM ($p = 0.002$). Post-hoc multiple comparisons showed that there were significant differences ($p<0.01$ for $T_{2w}^*$ and QSM; $p<0.05$ for $R_2^*$) between the uncorrected reconstruction and the other two reconstructions with corrections (i.e., MC or MC+Lin.$\Delta B_0$). The subjects unintentionally performed 0.3-4.4 mm translation and 0.2-3.6º rotation, under which more translation/rotation could induce more variations in $T_{2w}^*$ (TE=32 ms) and the calculated $R_2^*$ and QSM images (e.g., subjects 3~6). Without any correction, the reconstructed results exhibited large variations from the reference. With MC or MC+Lin.$\Delta B_0$, RMSE reduced but was still significant, especially in the $R_2^*$ and QSM images calculated with multiple $T_{2w}^*$ images. For example, there was maximum 1 Hz in $R_2^*$ and 4 ppb in QSM reduction in RMSE when MC+Lin.$\Delta B_0$ was used, as compared to using MC-only reconstruction (subject 5), while the reductions were minimal in other subjects. Across the six subjects, there were 1-6 Hz variations from the reference $R_2^*$ and 7-19 ppb from the reference QSM when reconstructed with MC or MC+Lin.$\Delta B_0$.

**Discussion**

In this study, a navigator-based, joint motion and $B_0$ shift detections technique was developed on a conventional 3D ME-GRE MRI sequence for QSM and quantitative $R_2^*$ measurement. A reconstruction method for joint retrospective motion and $B_0$ drift corrections was proposed, and the effectiveness was demonstrated by simulations and routine human brain scans. The results showed that efficient $B_0$ shift correction with conjugate-phase reconstruction can be performed on less than 10 clustered k-space segments.

Previous research demonstrated the effects of motion and temporal $B_0$ shift on the $T_{2w}^*$ image quality and attached great importance of the motion and/or temporal $B_0$ shift corrections on the $T_{2w}^*$ image (5-12). In our work, when the subject had 4 mm translational and 5º rotational motion, the $B_0$ shift could be as significant as 7 Hz (Fig. 6a-b), which was similar to the previous report at 7.0 T and suggested the effect of the $B_0$ shift cannot be ignored at 3.0 T (5,7,29). The artifacts in $T_{2w}^*$ image and the variations of $R_2^*$ and QSM also depend on the distribution of the motion traveling across the k-space from the center to the edge. Our results showed that even though the occurrence of motion was far away from the k-space center, which determines most image contrast (see Fig. 6), $\Delta B_0$ due to large motion (~5 mm) could introduce non-negligible variations in $R_2^*$ and QSM (Fig. 6c).

It was reported that uncorrected motion itself could bias regional distribution of QSM (13). In the proposed method, not only the motion effects, but also the effects of the temporal $B_0$ change on QSM were evaluated, in which the measured QSM in routine scans could vary 20 ppb from the ones reconstructed with the proposed method (Fig. 7). These results can be explained by the fact that QSM depends on the phase of the image. There was still significant variation in the QSM obtained with MC Lin.$\Delta B_0$ from the proposed method, suggesting that linear $\Delta B_0$ correction was insufficient but higher-order spherical harmonics in $B_0$ change should be considered. Measurements on six subjects showed that both motion and $B_0$ shift could produce 20 ppb derivations in QSM, which was induced by no more than 2 mm translational and 2º rotational motion at 3.0 T. These results are supported by the previous report in which 5 ppb variation on QSM was reduced with motion and $B_0$ drift corrections at 7.0 T (11). Although the QSM variations depend on subjects' motion and $B_0$ shift, our results suggest that such effects could not be ignored at 3.0 T. Correcting such variation should improve QSM evaluations in neurodegenerative studies (3,4).

We used a dual-echo stack-of-spirals trajectory as the navigators for motion and $\Delta B_0$ detections. Compared to echo-planar imaging (EPI)-based volumetric navigators, off-resonance only causes blurring in spiral imaging, rather than image distortion/voxel displacement that is common in EPI, and could compromise the quality of the navigators for motion and $B_0$ drift detections if such effects were not effectively corrected (5,11,21). Unlike EPI, spiral imaging also provides short TE readout, which helps reduce signal loss in navigator acquisitions (21). In this approach, movements in routine scans were captured with navigator acquisition interval of 1 s, and the navigator acquisition time was accelerated with parallel imaging. An additional feature of the design is that the 3D navigators are inserted between the RF excitations, which is independent of the GRE acquisition and thus provide the flexibility of changing TE and increasing echo numbers. In this way, traditional ME-GRE for quantitative $R_2^*$ and/or QSM can be combined with this method.

Various efficient reconstructions with joint motion and temporal $B_0$ shift corrections have been proposed (5,6). For example, $B_0$ shift was simplified by a linear spatial model across the whole brain, in which the low-order spherical harmonics approximation of $\Delta B_0$ was absorbed into the acquired signal model to facilitate efficient reconstruction algorithm with motion-induced k-space rotation and linear phase change (6). Although the effectiveness of using linear $B_0$ shift approximation was demonstrated to correct the effects due to respiration, the artifacts by $\Delta B_0$ induced by head motion could not be sufficiently reduced (5,29). More methods have been proposed to incorporate higher-order spherical harmonics in $B_0$ shift to improve the correction accuracy (5,7-9,12). Among these methods, the $B_0$ shifts were clustered and the corresponding the segmented k-space data were reconstructed with interpolation-based non-uniform fast Fourier transform (NUFFT) to achieve the computational efficiency (5). In this work, a simple approach was proposed using conjugate-phase reconstruction on the k-space data segmented by the clustered $B_0$ shift during acquisition. Since the reconstruction of each k-space cluster is independent of each other, it can be done in parallel during the whole iterations for each k-space segment. Furthermore, various non-uniform k-space reconstructions can be used for motion correction without any modification in our proposed method. Alternatively, our proposed method can be combined with prospective motion correction and thus the prospectively motion-corrected uniform k-space reconstruction can be utilized without iterations in parallel imaging (e.g., by GRAPPA) followed by conjugate-phase correction for $B_0$ shift, which is potentially more efficient than using iterations in the previous joint motion and $B_0$ shift corrections.

In conclusion, our proposed approach of using dual-echo 3D spiral navigators and a novel conjugate-phase reconstruction method can improve susceptibility and $R_2^*$ measurements, and it can be used as a valuable tool in studying iron deposition, demyelination, and other processes in neurological diseases.

**ACKNOWLEDGEMENTS**

The study was supported by the National Institutes of Health Grant/Award Numbers: R21AG064405, R01AG07093, R01AG072603, P30AG066511.

quantitative imaging of cortical $R_2^*$ and magnetic susceptibility at 0.3 mm in-plane resolution at 7 T. Neuroimage 2023;270:119992.

**Figure Legends**

**Figure 1**    Gradient-echo signal reconstruction diagram with motion and temporal $B_0$ shift corrections at TE. The k-space (k.) lines are clustered into $n$ segments (seg.) based on the $\Delta B_0$ for each k-space line. Reconstruction (recon.) with motion correction (MC) is performed with the translationally corrected k-space data and the rotated k-space, followed by phase ($\phi_n$) corrections for each segment and the summation of all the segmental reconstructions.

**Figure 2**    Numerical simulations for motion and temporal $B_0$ shift corrections. (a) The simulated Shepp-Logan head phantom with simulated 32-channel receiver sensitivity maps and $\Delta B_0$ due to motion; (b) simulated two-pixels translation, two-degrees rotations with linear $\Delta B_0$ in a 2D under-sampled k-space center (10 lines); (c) comparisons of the uncorrected, motion-corrected (MC), motion and linear (lin.) $\Delta B_0$ corrected, motion-corrected with conjugate (conj.) phase with $\Delta B_0$, and their magnitude ($\Delta$Mag.) and phase differences ($\Delta$Phase) between the simulated phantom (ground truth); (d) comparisons of the RMSE of different reconstructed image (mag. and phase) from the ground truth.

**Figure 3**    The ME-GRE sequence with 3D dual-echo spiral navigators.

**Figure 4**    Slices of the magnitude of the first echo (a) and the $B_0$ field map (b) from a dual-echo 3D spiral navigator.

**Figure 5**    Results from a subject with gradual movement. (a) The detected motion (rotation and translation (trans.)) and the averaged $\Delta B_0$ across the whole brain from the navigator magnitude and the phase respectively; (b) The RMSE of the $\Delta B_0$ relative to the centroids of n (1-20) clusters were calculated ($RMSE_n$) and the error $|RMSE_n\text{-}RMSE_{20}|/|RMSE_1\text{-}RMSE_{20}|$ were plotted with respect to the clusters (n) (top), and the clustering distribution (middle) and the clustered $\Delta B_0$ maps (bottom) are shown. (c) Comparisons of the reconstructed $T_{2w}^*$ (TE=32 ms), the calculated $R_2^*$, the QSM with different reconstructions, and the difference between linear (lin.) and conjugate (conj.) $\Delta B_0$ corrections in addition to motion correction (MC). The yellow box shows the zoomed-in images, and the yellow arrows show the artifacts.

**Figure 6**    Results from a subject with unintentional sudden movement. (a) The detected motion (rotation and translation (trans.)) and the averaged $\Delta B_0$ across the whole brain from the navigator magnitude and the phase respectively; (b) The RMSE of the $\Delta B_0$ relative to the centroids of n (1-20) clusters were calculated ($RMSE_n$) and the error $|RMSE_n\text{-}RMSE_{20}|/|RMSE_1\text{-}RMSE_{20}|$ were plotted with respect to the clusters (n) (top), and the clustering distribution (middle) and the clustered $\Delta B_0$ maps (bottom) are shown. (c) Comparisons of the reconstructed $T_{2w}^*$ (TE=32 ms),

the calculated R$_2$*, the QSM with different reconstructions, and the difference between linear (lin.) and conjugate (conj.) ΔB$_0$ corrections in addition to motion correction (MC). The yellow box shows the zoomed-in images, and the yellow arrows show the corrected artifacts.

**Figure 7** (a) The maximum translation (trans., mm) and rotation (rot., degree) detected from the navigators across the time points from six subjects. At each time point, the maximum translation (mm) and rotation are the largest value in the three translational dimensions and rotational orientations, respectively. (b-d) The RMSE results of T$_{2w}$* at TE = 32 ms (b), R$_2$* (c) and QSM (d) from between different reconstructions and the MC + Conj.ΔB$_0$ reconstruction.

**Figure S1** Reconstructed images (a) and NRMSE (b) with MC+Conj.ΔB$_0$ with different ΔB$_0$ clustering number. The yellow arrow in (a) showed the artifacts.

# Appendix

In the presence of B$_0$ shift due to motion, the term $E^\dagger E$ in Eq. [5] can be written as a $F^\dagger F + \Delta$, where $\Delta$ is a perturbed matrix of the term $F^\dagger F$ with the entry at location $(r_i, r_j)$:

$$\Delta(r_i, r_j) = \sum_{c=1}^{n_c} \sum_{n=1}^{n_k} B_{1,c}(r_i) B_{1,c}(r_j) \{ e^{k_n(r_i - r_j)} [e^{i \cdot \gamma \cdot \omega(r_i, r_j, k_n) \cdot TE} - 1] \}, \quad \text{[A.1]}$$

where $n_c$ is the receiver coil number, $n_k$ is the length of the k-space data points, and $\omega(r_i, r_j, k_n)$ is the B$_0$ difference between the location $r_i$ and $r_j$ while data are acquired at k-space location $k_n$

$$\omega(r_i, r_j, k_n) = \Delta B_{0,k_n}(r_i) - \Delta B_{0,k_n}(r_j). \quad \text{[A.2]}$$

Then, Eq. [5] can be rewritten as

$$\widehat{M} = (F^\dagger F + \Delta)^{-1} E^\dagger S'. \quad \text{[A.3]}$$

Using Power series expansion for $(F^\dagger F + \Delta)^{-1}$, Eq. [A.3] can be rewritten as

$$\widehat{M} = [(F^\dagger F)^{-1} + R] E^\dagger S', \quad \text{[A.4]}$$

where R is the higher orders of Power series expansion for $(F^\dagger F + \Delta)^{-1}$:

$$R = \sum_{i=1}^{\infty} (F^\dagger F)^{-1} [-\Delta (F^\dagger F)^{-1}]^i \quad \text{[A.5]}$$

**Figures**

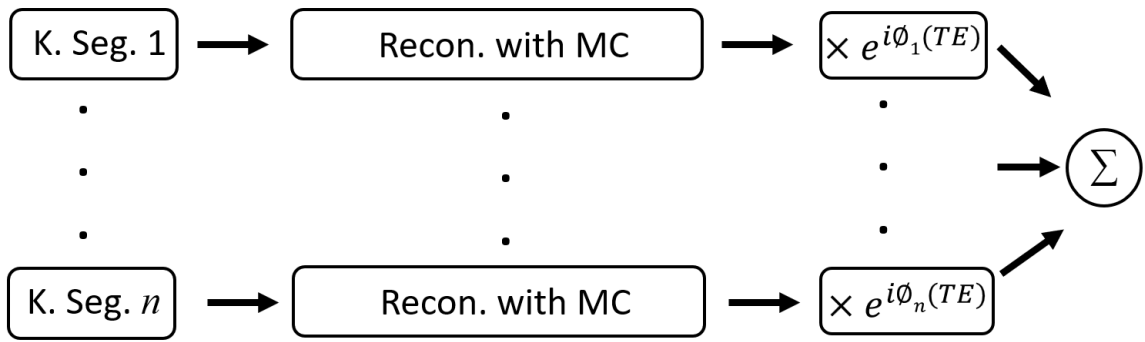

**Figure 1**   Gradient-echo signal reconstruction diagram with motion and temporal $B_0$ shift corrections at TE. The k-space (k.) lines are clustered into *n* segments (seg.) based on the $\Delta B_0$ for each k-space line. Reconstruction (recon.) with motion correction (MC) is performed with the translationally corrected k-space data and the rotated k-space, followed by phase ($\phi_n$) corrections for each segment and the summation of all the segmental reconstructions.

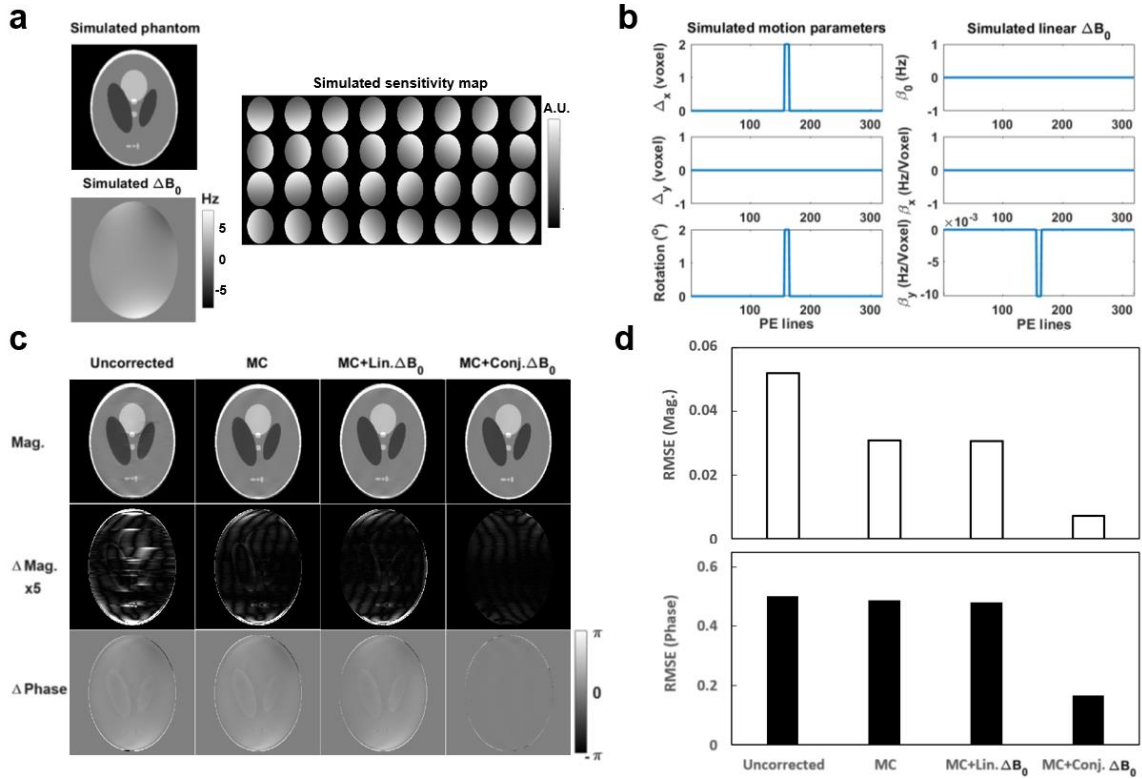

**Figure 2** Numerical simulations for motion and temporal $B_0$ shift corrections. (a) The simulated Shepp-Logan head phantom with simulated 32-channel receiver sensitivity maps and $\Delta B_0$ due to motion; (b) simulated two-pixels translation, two-degrees rotations with linear $\Delta B_0$ in a 2D under-sampled k-space center (10 lines); (c) comparisons of the uncorrected, motion-corrected (MC), motion and linear (lin.) $\Delta B_0$ corrected, motion-corrected with conjugate (conj.) phase with $\Delta B_0$, and their magnitude ($\Delta$Mag.) and phase differences ($\Delta$Phase) between the simulated phantom (ground truth); (d) comparisons of the RMSE of different reconstructed image (mag. and phase) from the ground truth.

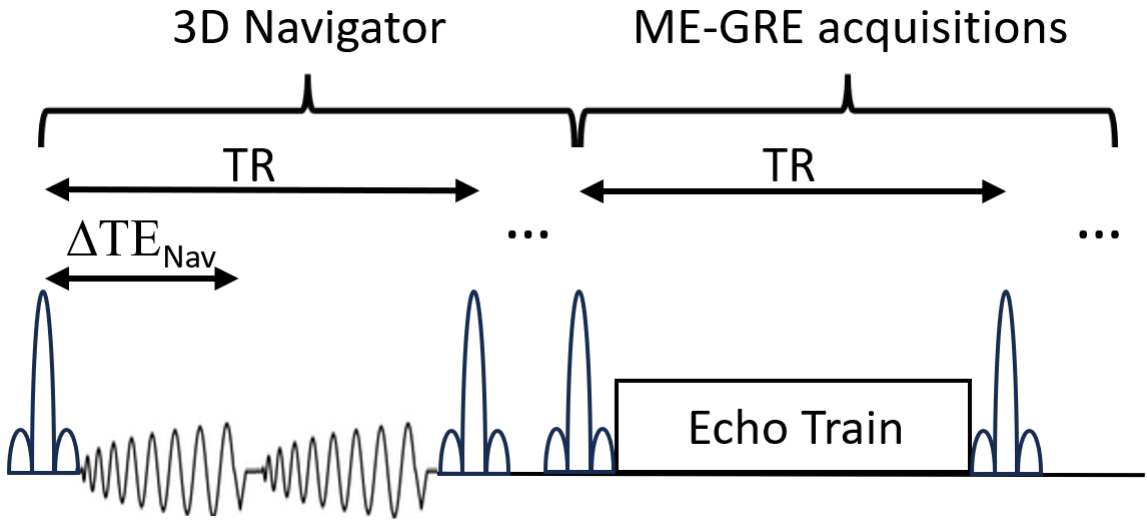

**Figure 3** The ME-GRE sequence with 3D dual-echo spiral navigators.

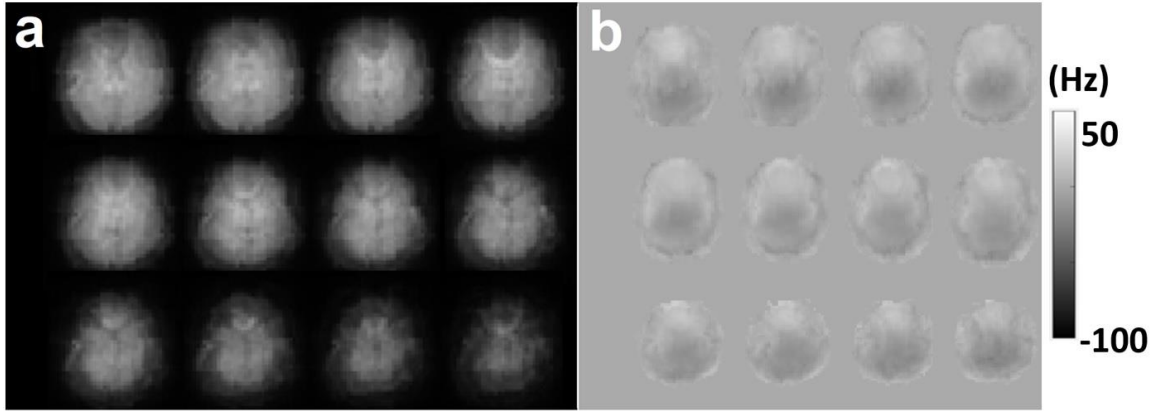

**Figure 4**  Slices of the magnitude of the first echo (a) and the $B_0$ field map (b) from a dual-echo 3D spiral navigator.

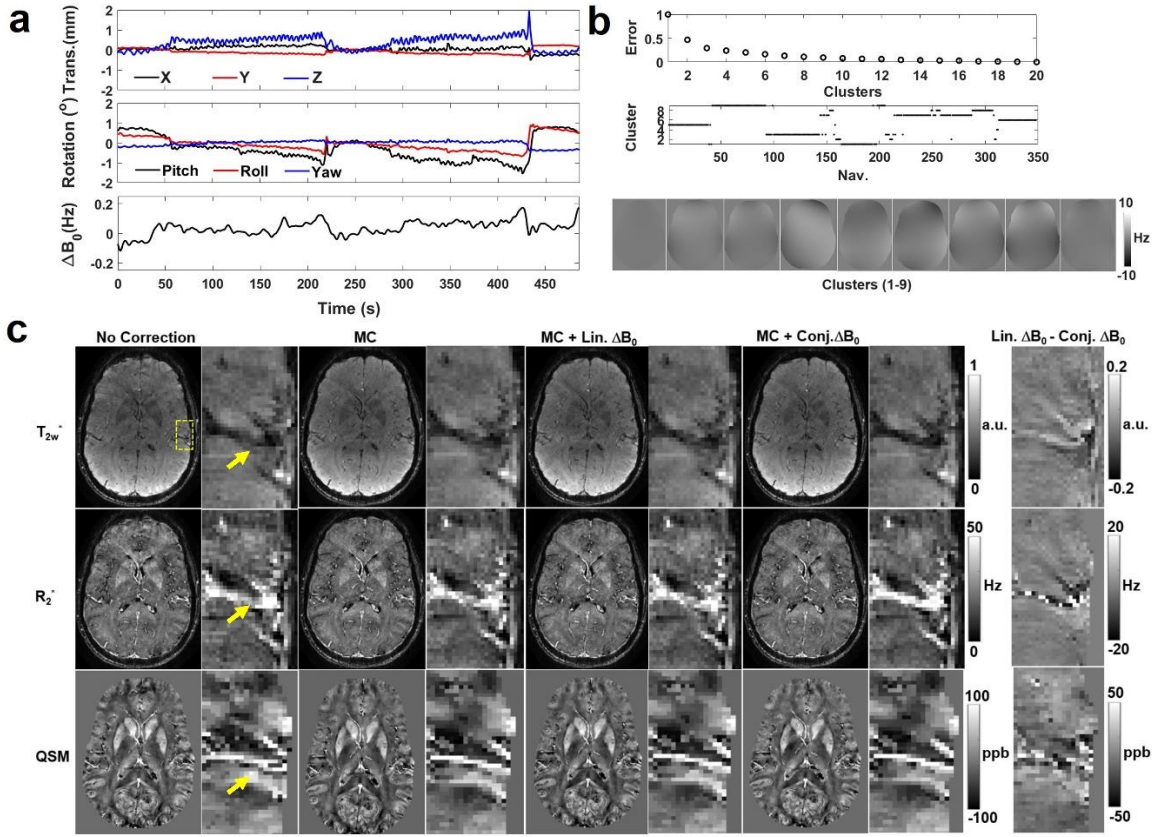

**Figure 5** Results from a subject with gradual movement. (a) The detected motion (rotation and translation (trans.)) and the averaged $\Delta B_0$ across the whole brain from the navigator magnitude and the phase respectively; (b) The RMSE of the $\Delta B_0$ relative to the centroids of n (1-20) clusters were calculated ($RMSE_n$) and the error $|RMSE_n\text{-}RMSE_{20}|/|RMSE_1\text{-}RMSE_{20}|$ were plotted with respect to the clusters (n) (top), and the clustering distribution (middle) and the clustered $\Delta B_0$ maps (bottom) are shown. (c) Comparisons of the reconstructed $T_{2w}^*$ (TE=32 ms), the calculated $R_2^*$, the QSM with different reconstructions, and the difference between linear (lin.) and conjugate (conj.) $\Delta B_0$ corrections in addition to motion correction (MC). The yellow box shows the zoomed-in images, and the yellow arrows show the artifacts.

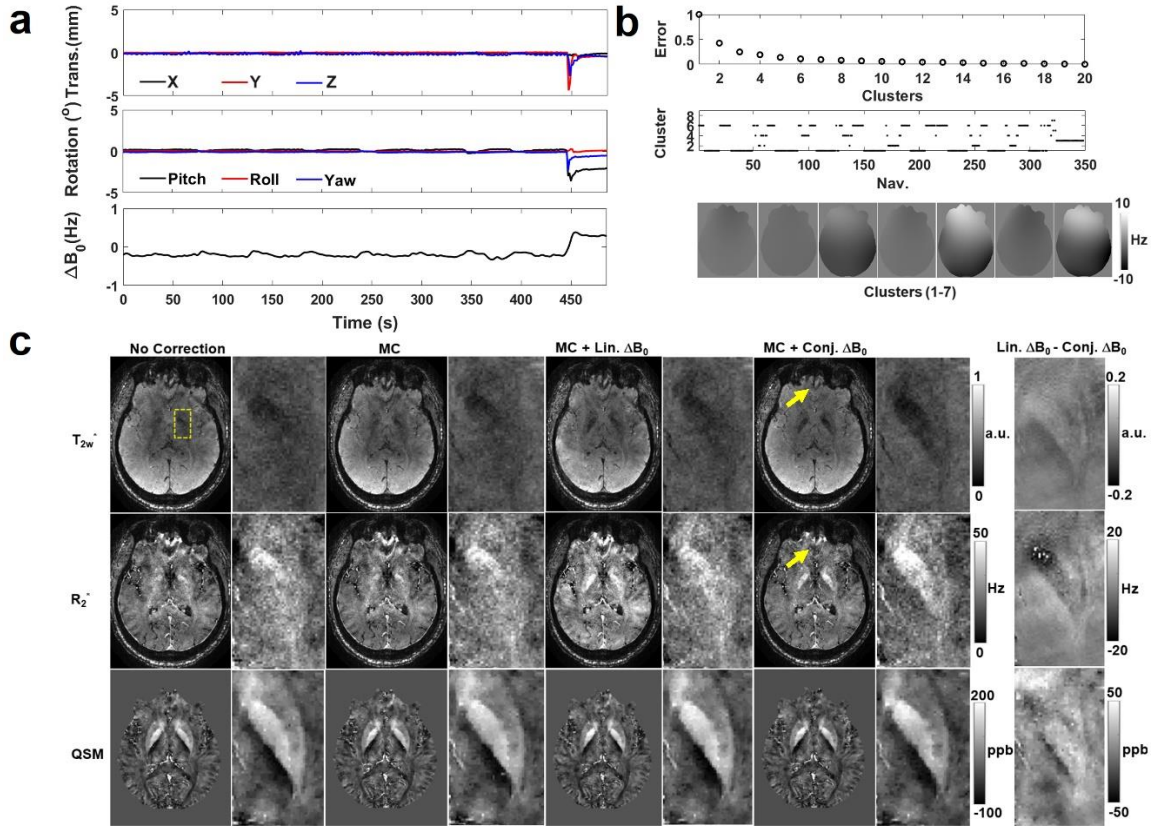

**Figure 6**    Results from a subject with unintentional sudden movement. (a) The detected motion (rotation and translation (trans.)) and the averaged $\Delta B_0$ across the whole brain from the navigator magnitude and the phase respectively; (b) The RMSE of the $\Delta B_0$ relative to the centroids of n (1-20) clusters were calculated ($RMSE_n$) and the error $|RMSE_n-RMSE_{20}|/|RMSE_1-RMSE_{20}|$ were plotted with respect to the clusters (n) (top), and the clustering distribution (middle) and the clustered $\Delta B_0$ maps (bottom) are shown. (c) Comparisons of the reconstructed $T_{2w}^*$ (TE=32 ms), the calculated $R_2^*$, the QSM with different reconstructions, and the difference between linear (lin.) and conjugate (conj.) $\Delta B_0$ corrections in addition to motion correction (MC). The yellow box shows the zoomed-in images, and the yellow arrows show the corrected artifacts.

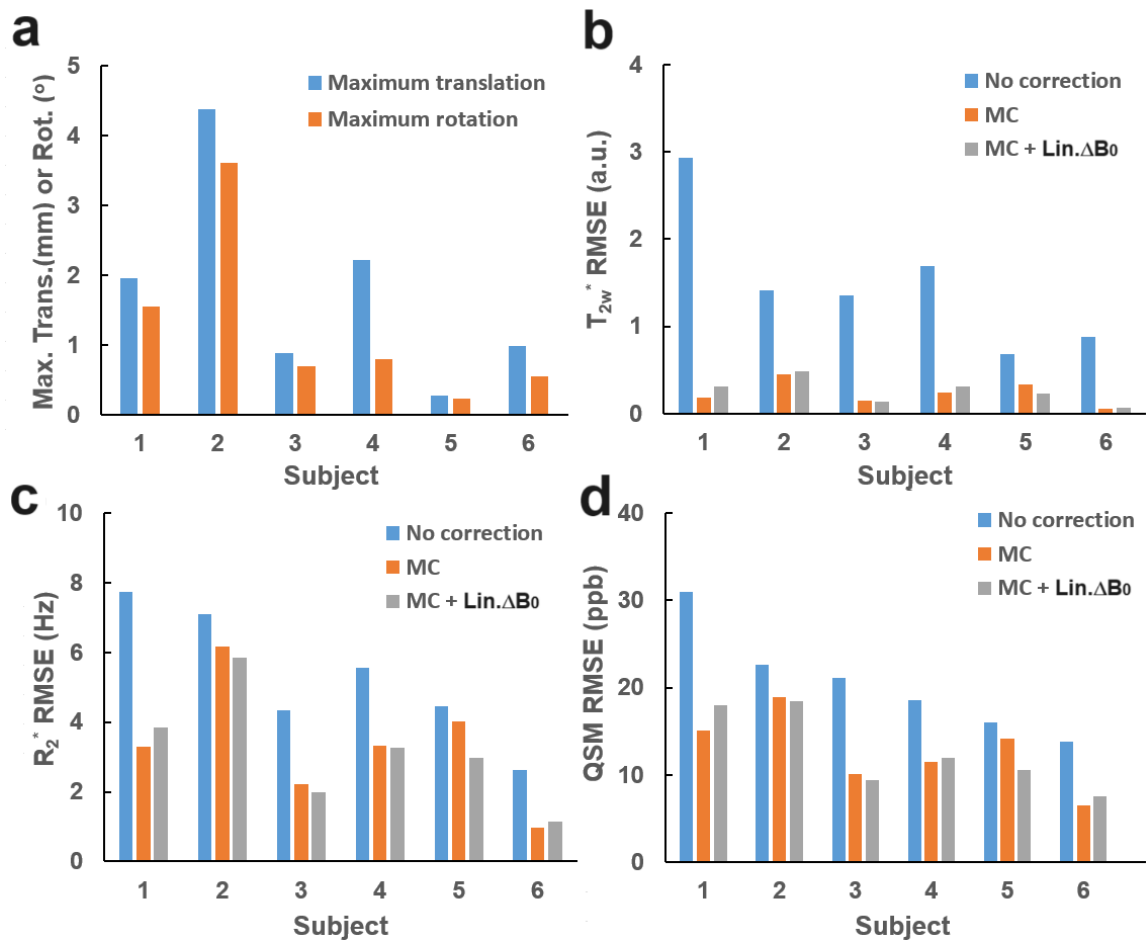

**Figure 7** (a) The maximum translation (trans., mm) and rotation (rot., degree) detected from the navigators across the time points from six subjects. At each time point, the maximum translation (mm) and rotation are the largest value in the three translational dimensions and rotational orientations, respectively. (b-d) The RMSE results of $T_{2w}^*$ at TE = 32 ms (b), $R_2^*$ (c) and QSM (d) from between different reconstructions and the MC + Conj.$\Delta B_0$ reconstruction.

# Supplemental materials for "Motion and temporal $B_0$ shift corrections for quantitative susceptibility mapping (QSM) and $R_2$* mapping using dual-echo spiral navigators and conjugate-phase reconstruction"

Yuguang Meng, Jason W. Allen, Vahid Khalilzad Sharghi and Deqiang Qiu

## Examine the correction accuracy with MC+Conj.$\Delta B_0$ with different $\Delta B_0$ clustering number

To examine the sufficiency of clustering number for $B_0$ shift correction, NRMSE of $T_{2w}^*$ images at TE=32 ms between the reconstructions with clusters from 1 to 20 and the maximum 20 clusters were calculated from a subject (Fig. 6) with significant motion. The reconstructed images and NRMSE were shown in Fig. S1.

## Supplemental Figures

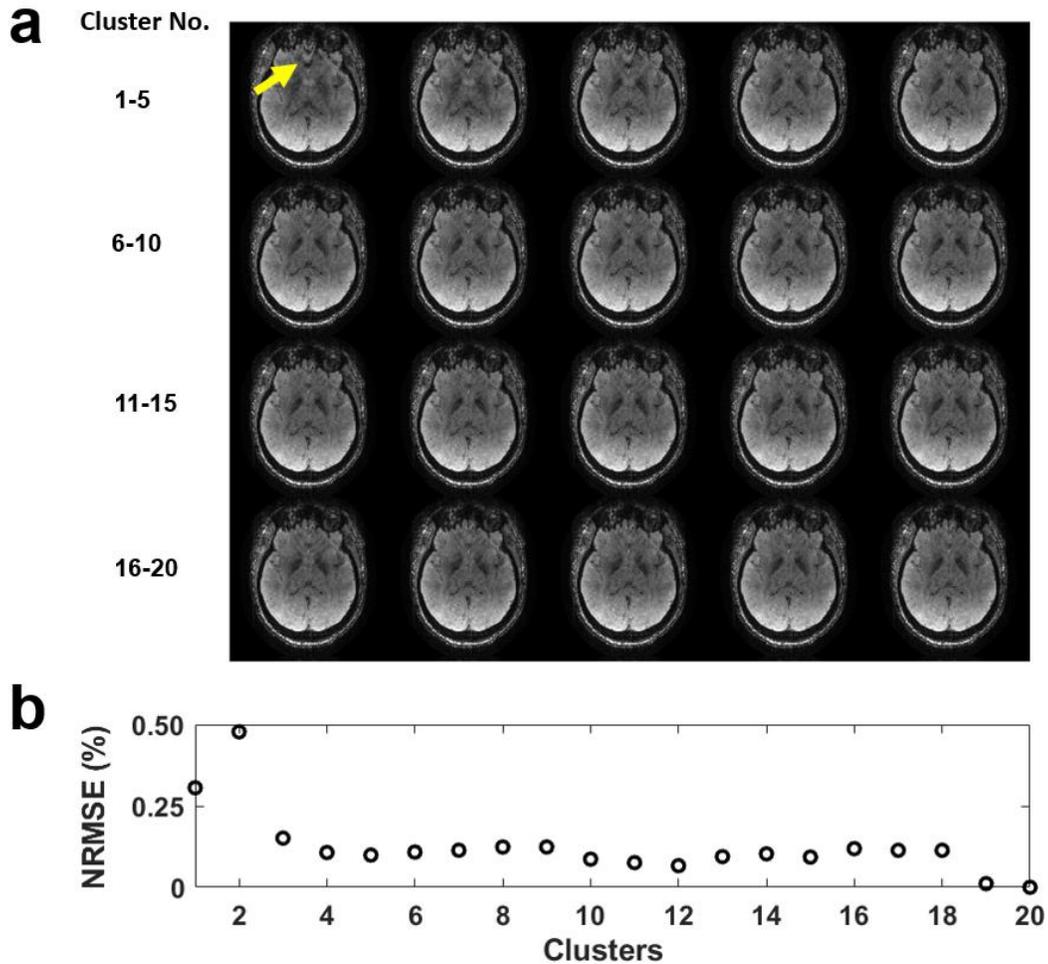

**Figure S1**   Reconstructed images (a) and NRMSE (b) with MC+Conj.$\Delta B_0$ with different $\Delta B_0$ clustering number. The yellow arrow in (a) showed the artifacts.